# The method of isolines for the shower axis determination

Kirov, I.N., Stamenov, J.N.

Institute for Nuclear Research and Nuclear Energy, 1784, Sofia, Bulgaria

The method of isolines (contours) is proposed to be used for determination of core location of extensive air shower on the array plane. As variable z=f(x, y) are used flux densities of particles, registered from ground detectors. Obtained isolines are approximated with circles, which centres show impact point of shower axis. The method was demonstrated on experimental data from Tien-Shan shower array in period 1974-82. The possibilities and difficulties of the method are discussed.

Introduction. The method of isolines (contours) is very often use when input data is collected in an irregular fashion, it is said to be scattered, irregular or random. Scattered data arise in many fields and problems such as: Earth Sciences: Geography, Geology, Geodesy, Hydrology and Geophysics; Meteorology: Weather measurements; Engineering: Computer aided geometric design; Medicine: Image reconstruction from samples.

The construction of isolines involve two stages: interpolation, which produces a regularly spaced data array from irregularly spaced data, so called gridding, and determination of contours as lines with constant values of depending variable.

The scattered data interpolation problem can be defined as: Given a set of $n$ irregularly distributed points

$$P_i = (x_i, y_i), i = 1,...,n$$

and scalar values $z_i$ associated with each point satisfying $z_i = z_i(x_i, y_i)$ for some underlying function $z(x,y)$, look for an interpolating function $\check{z} \sim \check{z}(x,y)$ such that for $i = 1,...,n$

$$\check{z} = z_i,$$

with assumption that all points $P_i$ (sometimes referred as nodes or mesh points) are distinct and that all the points are not collinear. This formulation concentrates in the case where the scattered points $P_i$ are on the plane, making the function $\check{z}$ a bivariate function that defines a surface in 3D space.

There is a plenty of interpolation methods and it is not a unique way to classify them. In general they could be global, where in the interpolation are used all the nodes, and local, where only nodes in some circle around the interpolation point are used.

The calculation of isoline points usually is based on the linear interpolation between neighboring grid points.

The next step of the method is the fitting simple contours (circular arcs) to the obtained isoline points. The need of approximating scattered points by a circle or a circular arc arises in physics, biology and medicine, archeology, industry, computer graphics, coordinate metrology, etc. There are various methods and strategies for curve fitting, e.g. Hough transform, moment method, least squares method, etc.

The least squares fit of circles is based on minimizing the mean square distance from the circle to the data points. Given $n$ points $(x_i, y_i)$, $1 < I < n$, the objective function is defined by

$$F = \sum d_i^2$$

where $d_i$ is the geometric distance from the point $(x_i, y_i)$ to the circle. If the circle satisfies the equation

$$(x - a)^2 + (y - b)^2 = R^2$$

where $(a, b)$ is its center and $R$ its radius, then

$$d_i = \sqrt{(x_i-a)^2 + (y_i-b)^2} - R .$$

The algorithms that use this objective function (to minimize the sum of the squares of the distances to the given points) are called geometric. The so called algebraic fitting minimizes the square sum of the algebraic function values at each given point:

$$F_a = \sum((x_i-a)^2 + (y_i-b)^2 - R^2)^2 .$$

Method. The proposed method involve two stages:
1. construction of isolines with constant particle densities, and
2. approximation of obtained isolines with circles, which centers give the core location of the shower.

As a preparing step before calculations we transform detectors coordinates in a plane, perpendicular to the axis direction. This is realized by means of two rotations on axis direction angles - azimuth angle $\varphi$ and zenith angle $\theta$.

We carried out an averaging of the data to exclude relatively big fluctuations of the data. This averaging performs the role of the smoothness and approximation actions. After that we divided averaging densities into some groups by ascending order of density values. The number of groups gives a number of asked isolines and the average of the group gives the constant value of isoline.

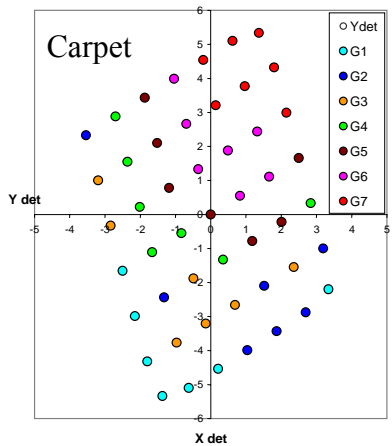

Fig. 1. Arrange in 7 group by ascending order of averaging particle densities for the central part ("carpet") of Tien-Shan EAS array.

The coordinates of the isoline points are calculated by linear interpolation between vertexes of every tetragon of the averaging grid.[ Note that in all operations we essentially take advantage from specific array geometry.]

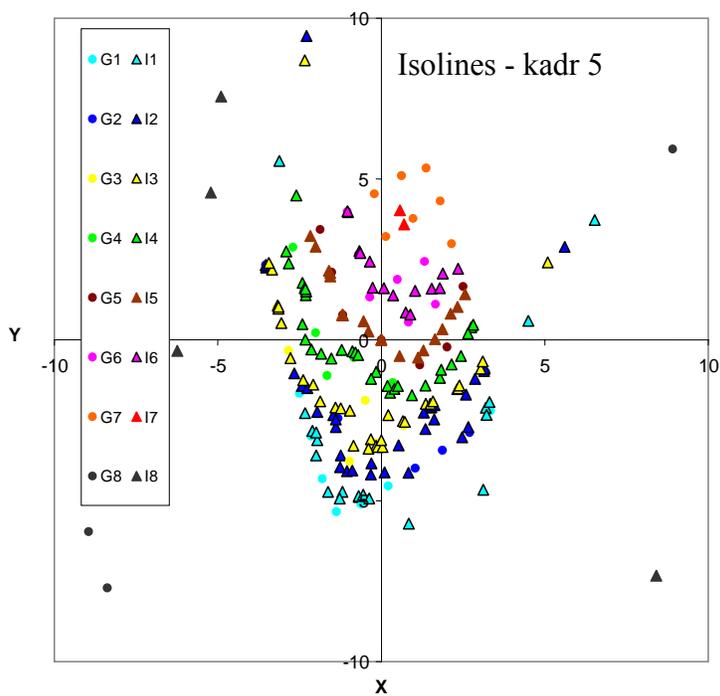

*Fig. 2. Isolines, obtained for average densities of 7 groups of central part ("carpet") and 4 points at 15 m. and 4 points of 20 m. from the array centre.*

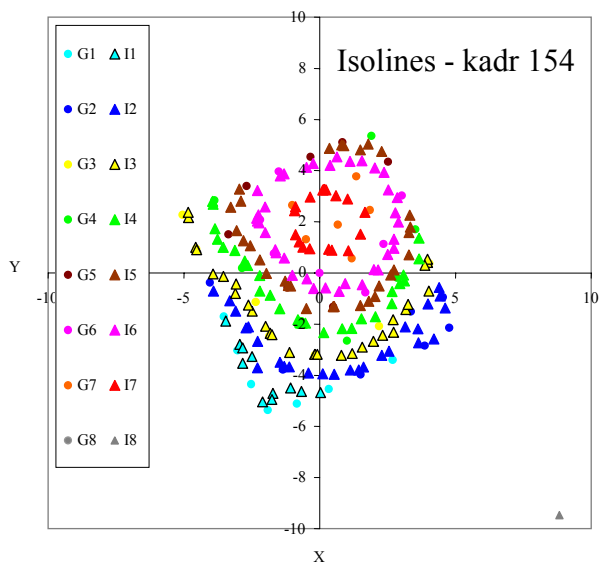

*Fig. 3. Isolines for a "central" event The axis is located in the central part of array.*

The points of every isoline are fitted with the circle. We use two fitting algorithms: modified least square method [2] and iterative weighted inversion method [3].

In inversion method is used a property of conformal mappings that circles through the origin map to straight lines under inversion. Therefore we can use a standard straight line fit formula (total least squares) in new coordinates. Taking the inverse of the fitted straight line using the same pole of inversion we retrieve the fitted circle.

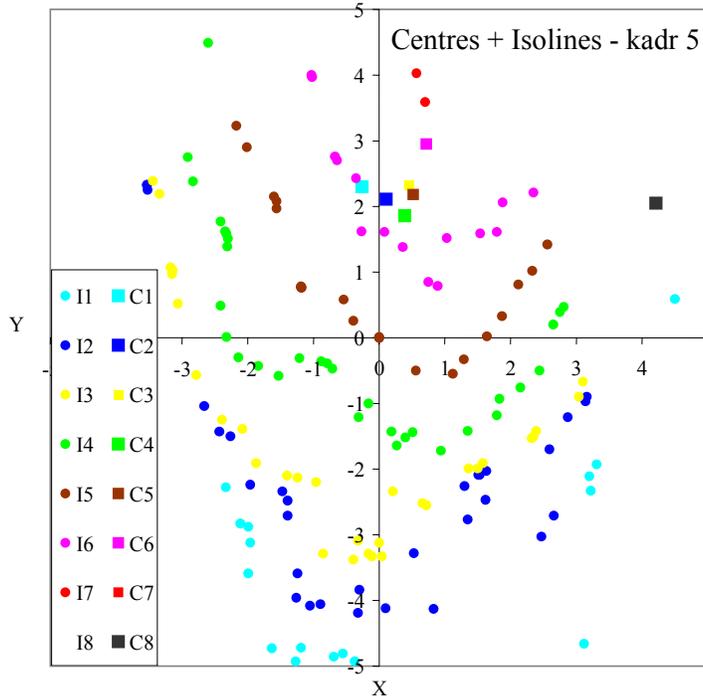

*Fig. 4. The centres of obtained circles [in squares] for corresponding isolines.*

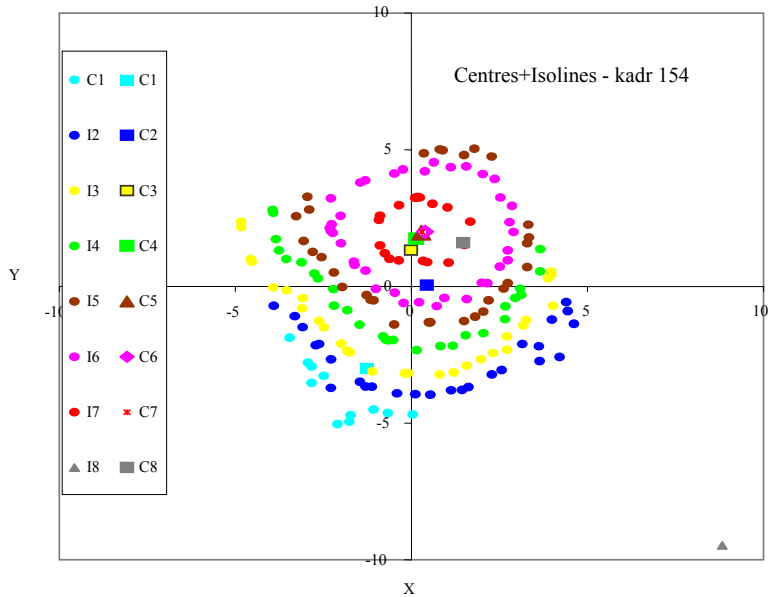

*Fig. 5. Centers of obtained circles for corresponding isolines for central event.*

Discussion.

The main advantages of proposed method are:
1. The method is free of any apriority given function of density distribution and is not needed of initial values of shower size and age parameter.
2. Isolines give a possibility to investigate lateral distribution on a plane in more details.
3. It can perform as good initial value for the shower parameters determination.

The disadvantages and difficulties are
1. to find the most appropriated interpolation algorithm. For example, you can see on Fig. 6 directly enclosure interpolation method – in gray color are the regions with negative densities - that is nonsense.

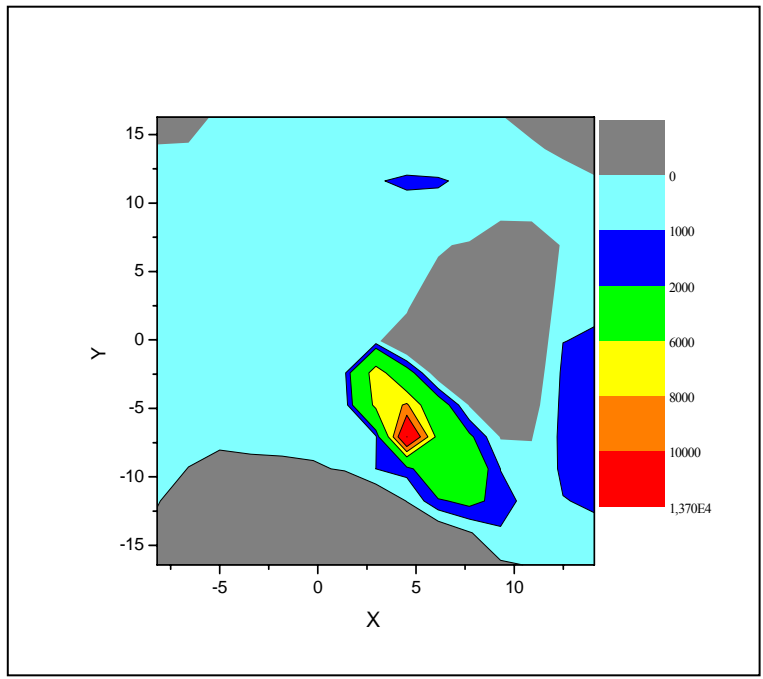

*Fig. 6. Contour map of one event, obtained with Origin by the approximation of Renka-Cline.*

2. it may be better to approximate isolines with planar curves of second order and from their parameters ( eccentricity, etc.) to estimate the curve type.

Acknowledgements.  We wish to express our thanks to Cosmic Ray Department of Lebedev Physical Institute to place experimental data to our disposal. I. Kirov thanks to Prof. N. Angelov from JINR-Dubna for suggested idea many years ago.

We thank IT and BEO-Mussala teams for the technical support. This work was partially supported from FP6 project BEOBAL.